\documentclass[fleqn,twoside]{article}
\usepackage{espcrc2}

\usepackage{graphicx}
\usepackage[figuresright]{rotating}

\newcommand{\AmS}{{\protect\the\textfont2
  A\kern-.1667em\lower.5ex\hbox{M}\kern-.125emS}}

\vspace{-5pc}
\title{Excitations of the torelon}

\author{K. J. Juge\address{Institute of Theoretical Physics, University of Bern, Sidlerstrasse 5, CH-3012 Bern, Switzerland}
        J. Kuti\address{Department of Physics, University of California at San Diego, La Jolla, USA 92093-0319},
        F. Maresca\address[TCD]{School of Mathematics, Trinity College, 
Dublin 2, Ireland}\thanks{Talk presented by F. Maresca},
        C. Morningstar\address{Department of Physics, Carnegie Mellon University, Pittsburgh, PA, USA 15213-3890}
        and
        M. Peardon\addressmark[TCD].}
       
\begin{document}

\begin{abstract}
The excitations of gluonic flux tube in a periodic lattice are
examined. Monte Carlo simulations from an anisotropic lattice are
presented and the comparison with effective string models is
discussed. 
\end{abstract}

\maketitle

\section{INTRODUCTION}
It is believed that the confining regime of Yang-Mills theory may be 
described by some kind of effective string model. The energy of the
flux tube between static quark sources in 4d SU(3) has been studied in
detail \cite{JKM1,JKM2}:
the grouping of excited states into bands and the corresponding 
energy gap dependence on $r$ at large distances is suggestive of the string 
picture, but there is a significant fine structure.
Recent studies in 3d Z(2), SU(2) and compact U(1) \cite{CM} confirmed a string
formation at large separation with less pronounced fine structure.
In the following calculations we analyse the spectrum of the QCD periodic flux 
tube. The torelon has no fixed colour sources and so provides a particularly 
favourable theoretical environment in which to observe the onset of string
behaviour \cite{Teper1}.  

\section{SIMULATION DETAILS \label{sec:simdetail}}
On the lattice, torelons may be created using colour-singlet traces of ordered product of link 
matrices encircling a spatial lattice direction.
Excitations are obtained by projection onto longitudinal momentum $p_z$
eigenstates and onto different irreps of rotations in the plane transverse to 
the flux. The energy spectrum of a QCD flux tube is then estimated by 
Monte-Carlo measurement of the correlation function, 
\begin{equation}
 C(t) = \langle \phi^{\dagger}(t) \phi(0) \rangle \stackrel{t
\rightarrow \infty}{\rightarrow} | \langle\mbox{vac}| \phi | 0 \rangle
|^2 e^{-E_{0}t},
\end{equation}
where $E_{0}$ is the energy of the lightest state which can be created
by the operator $\phi(t)$.
Because the signal decreases exponentially as t is increased, it is very 
important to obtain the asymptotic behaviour of the correlator as quickly as 
possible.  In order to do so we choose operators for which the overlap with 
the lightest state is as large as possible.
We use an anisotropic lattice in which the
temporal spacing $a_t$ is much smaller than that in the spatial directions
$a_s$ to exploit the enhanced signal-to-noise of the correlation function
at smaller temporal separations.
The center symmetry of the Lagrangian 
allows us to restrict our torelon operators to be positive under charge
conjugation. 
The lattice symmetry of rotations about the winding axis $z$, is 
$C_{4\nu} \otimes Z(\cal{R})$ for $p_z=0$ or $C_{4\nu}$ for $p_z \ne 0$.
$C_{4 \nu}$ contains rotations of $\frac{\pi}{2}$ and 
the reflection in the $xy$ plane ($\cal{P}$-parity) and has five irreducible
representations: $A_1$,$A_2$,$B_1$,$B_2$ and $E$.
$Z(\cal{R})$ denotes the two-element group consisting of the
identity operation and the reflection about the midpoint on the principal axis
$z$ ($\cal{R}$-parity). States that are even/odd under $\cal{R}$-parity
are labeled by the subscripts $g$/$u$ respectively.

\begin{figure}
{\resizebox{!}{55mm}{\includegraphics*{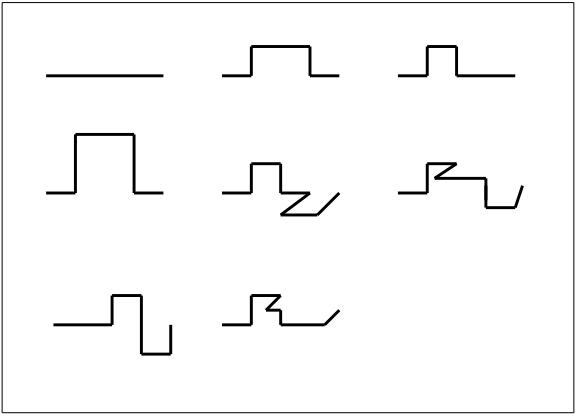}}}
\vspace{-8ex}
\caption{The eight prototype lattice paths used in the construction of torelon
operators in this work.
\label{fig:ops} }
\vspace{-2ex}
\end{figure}

Simulations were performed on three different lattices of extent in the $z$ 
direction of 8, 12 and 16 $a_s$. The lattice action was Symanzik improved and
included a two-plaquette term, designed to reduce cut-off artefacts in the
scalar glueball mass. The anisotropy was set to $a_s/a_t=6$ with $a_s=0.21$ fm 
(from $r_0$). Linear combinations of the eight operators shown in 
Fig.~\ref{fig:ops} transforming irreducibly under the lattice symmetries were 
computed on two smeared sets of links and a variational methods was employed 
on the 16 operators to optimise overlaps. 
\begin{table}[ht]
\caption{Lowest string energy levels and their corresponding string and QCD
states. The operator $a_m^{s(d) \dagger}$ creates an m string mode of
right(left) chirality. 
\label{table:1}}
\begin{tabular}{lll}
\hline
\mbox{Level} & \mbox{State}& \mbox{QCD $L^P_R(p_z)$}\\
\hline
0 & $|0\rangle$   & $\Sigma_g^+(0)$ \\
\hline
1 & $(a_{1}^{s \dagger}+a_{1}^{d \dagger} , a_{1}^{s
\dagger}-a_{1}^{d \dagger})|0\rangle$ & $\Pi(1)$ \\
\hline
2 & $(a_{1}^{s \dagger} \tilde{a}_{1}^{s \dagger} + a_{1}^{d
\dagger} \tilde{a}_{1}^{d \dagger})|0\rangle$ 
          & $\Delta_g(0)$ \\
 & $(a_{1}^{s \dagger} \tilde{a}_{1}^{s \dagger} - a_{1}^{d \dagger}
\tilde{a}_{1}^{d \dagger})|0\rangle$
          & $\Delta_g(0) $ \\
& $(a_{1}^{d \dagger}\tilde{a}_{1}^{s \dagger} - a_{1}^{s \dagger}
\tilde{a}_{1}^{d\dagger})|0\rangle$ 
          & $\Sigma_u^-(0)$ \\
&$(a_{1}^{d \dagger}\tilde{a}_{1}^{s \dagger} + a_{1}^{s \dagger}
\tilde{a}_{1}^{d\dagger})|0\rangle$ 
          & $\Sigma_g^{+*}(0)$ \\ 
\cline{2-3}
&  $((a_{1}^{s \dagger})^2 + (a_{1}^{d \dagger})^2)|0\rangle$
          & $\Delta(2)$ \\
&  $((a_{1}^{s \dagger})^2 - (a_{1}^{d \dagger})^2)|0\rangle$
          & $\Delta(2) $ \\
&  $a_{1}^{d \dagger} a_{1}^{s \dagger}|0\rangle$
          & $\Sigma^+(2)$ \\
&  $(a_{2}^{s \dagger} + a_{2}^{d \dagger} ,a_{2}^{s \dagger} -
a_{2}^{d \dagger})|0\rangle$
          & $\Pi(2)$ \\
\hline
3\footnotemark[1]
 & $(\tilde{a}_{1}^{ s \dagger} a_{2}^{s \dagger} + \tilde{a}_{1}^{ d
\dagger} a_{2}^{d \dagger})|0\rangle$
          & $\Delta(1)$ \\
& $(\tilde{a}_{1}^{ s \dagger} a_{2}^{s \dagger} - \tilde{a}_{1}^{ d
\dagger} a_{2}^{d \dagger})|0\rangle$
          & $\Delta(1)$ \\
& $(\tilde{a}_{1}^{ d\dagger} a_{2}^{s \dagger} + \tilde{a}_{1}^{ s
\dagger} a_{2}^{d\dagger})|0\rangle$
& $\Sigma^+(1)$ \\
\hline
\end{tabular}\\[2pt]
\footnotesize{\footnotemark[1]This level includes also the states
$0^-$ and
$1^{\star}$ with p=1 and states with p=3 not considered in
 our calculations}
\vspace{-0.05in}
\end{table}

\section{EFFECTIVE STRING THEORY}

For long QCD torelons, the two transverse oscillations of the tube
are the only degrees of freedom forced to be low-lying Goldstone modes. These 
are the degrees of freedom of an oscillating
string. In $d<26$ however, quantisation spoils Lorentz invariance and 
the string picture must then be regarded as an effective theory.

An important question is whether or not sufficiently long QCD torelons can be
well described by an effective string theory, such as the Nambu-Goto string in 
$d<26$, or the Polchinski-Strominger proposal \cite{PS}. Note that the simplest 
theories do not naturally include a number of effects that might also be 
relevant, such as rigidity terms or interactions with the QCD bulk.

At lowest order in an expansion about $1/L$, all these pictures give a 
common spectrum of states with energy gaps $2\pi N/L$. The quantum numbers of
some of these excitations for states up to $N=3$ are given in Table 
\ref{table:1}. This 
table is restricted to those states considered in the QCD calculation. This 
spectrum can be computed for the Nambu-Goto action and a computation of the 
$1/L^3$ corrections in the Polchinski-Strominger scheme is also under
investigation. The picture emerging from the analysis of these simple classes 
of effective string theories is that the fine structure is predominantly 
dependent on the longitudinal momentum along the string. 

\section{PRELIMINARY RESULTS}

\begin{figure}[htb]
\vspace{-4pc}
\rotatebox{270}{\resizebox{!}{85mm}{\includegraphics*{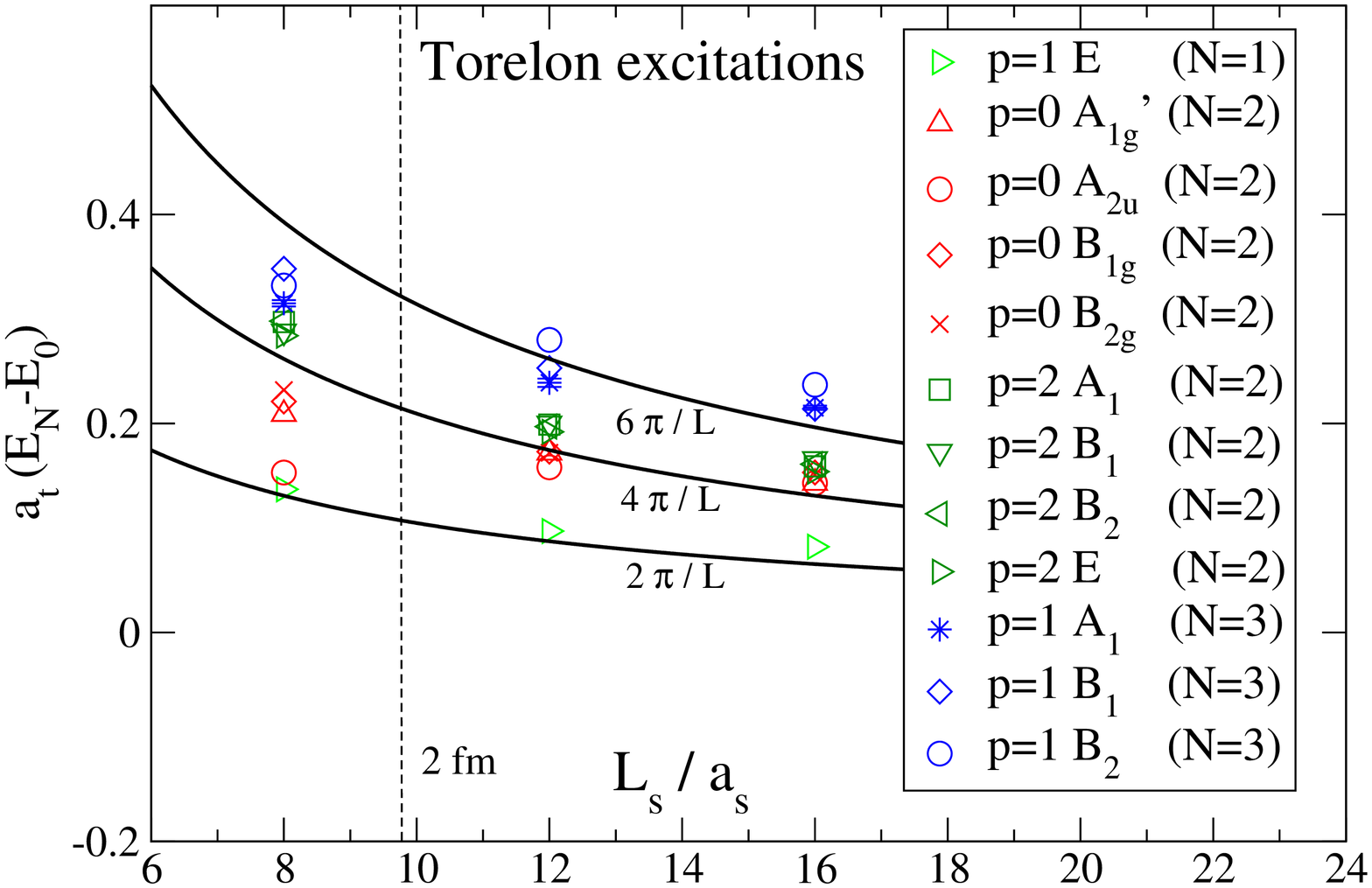}}}
\vspace{-2pc}

\rotatebox{270}{\resizebox{!}{85mm}{\includegraphics*{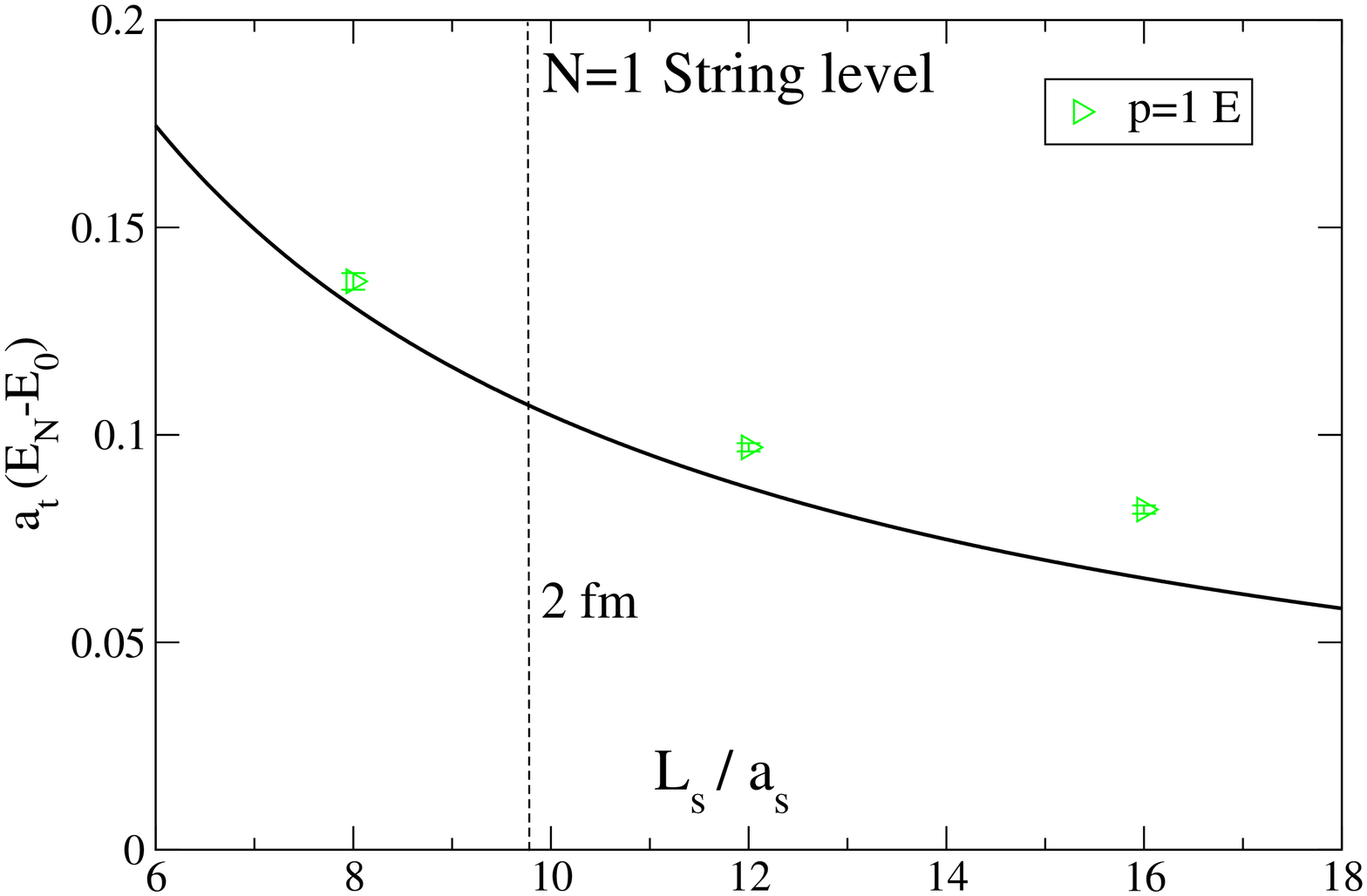}}}
\vspace{-2pc}

\rotatebox{270}{\resizebox{!}{85mm}{\includegraphics*{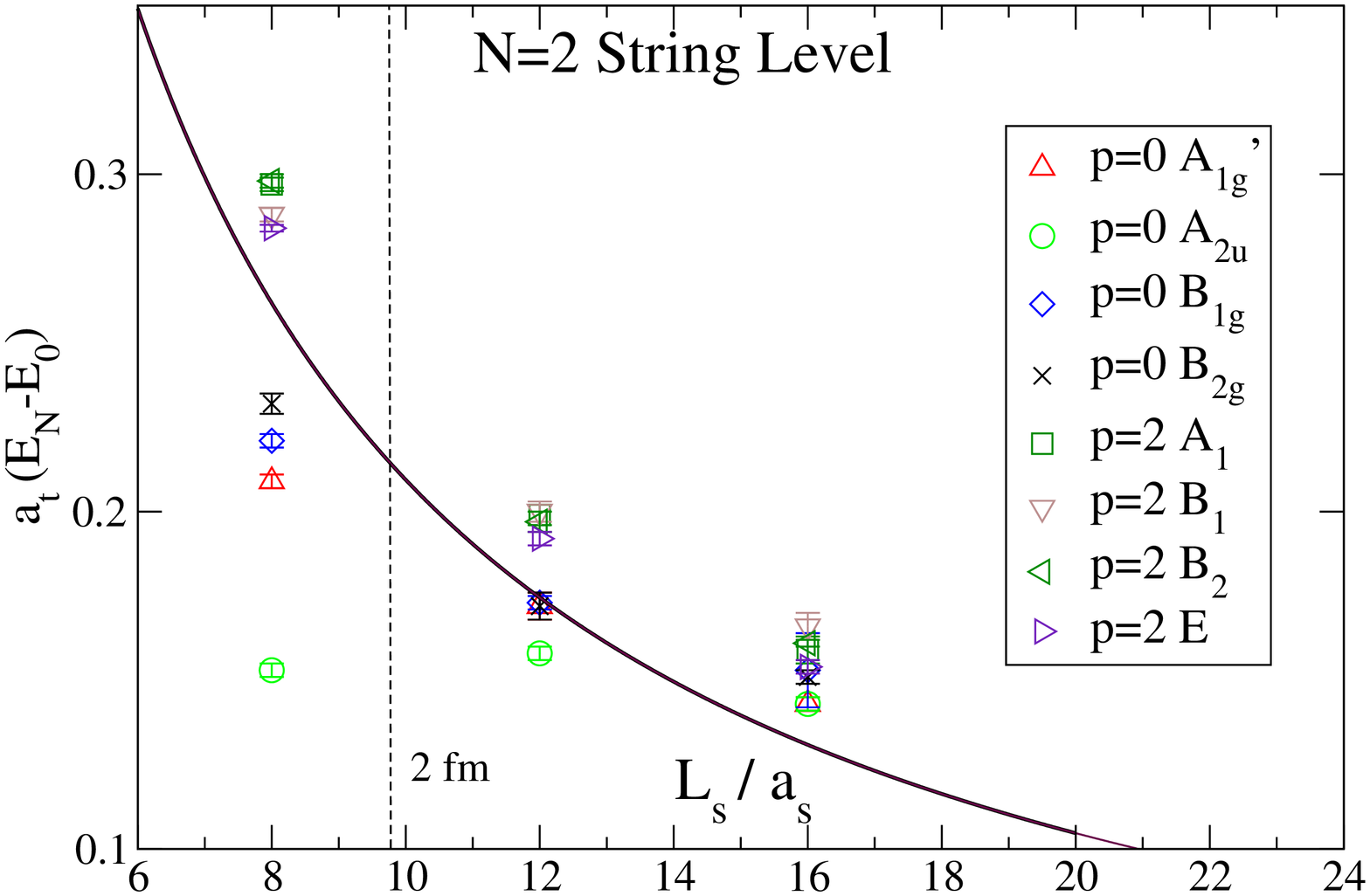}}}
\vspace{-3pc}

\caption{Energies of torelons of different lengths. 
The solid lines are the first order string prediction, $2\pi N/L$.}
\label{fig:Tor}
\end{figure}

The results from the three simulations described in Sec.~\ref{sec:simdetail}
are presented in Fig.~\ref{fig:Tor}. The value $L=2$ fm (determined from $r_0$)
is indicated by a vertical line. As can be seen in the first figure, all states
are correctly ordered according to their string classification, $N$ once $L>
2$ fm. 
The energy gap of all levels is qualitatively in agreement with the
first order string prediction $2\pi N / L $ as shown in the last figure. 

The second figure shows the $N=1$ level is close to the lowest-order string 
prediction, although deviations are resolved. The difference between the 
lowest-order string model and this excitation appears to grow as L is 
increased. 

In the $N=2$ states, significant structure is seen below 2 fm, while an 
extremely interesting pattern of degeneracies emerges at $L/a_s = 12$ and $16$. 
Here, the $p_z=0$ and $p_z=2$ states form two degenerate clusters. This is in 
agreement with indications from the effective string calculations that the 
fine structure is dependent on $p_z$ alone. The splitting between these
clusters decreases as $L$ increases. 

At $N=3$, for $L>2$ fm, there appears to be a single degenerate set (notice
from Table.~\ref{table:1} all 3
states have $p_z=1$) but a 
clear split between the $B_1$ and $B_2$ states is found. In the continuum 
these two states are expected to be degenerate and tests on possible lattice 
artifacts are under investigation. Extra operators for these states are 
being studied.

\section{CONCLUSIONS}

In this work, we have demonstrated that high-resolution measurements of the 
QCD torelon and its excitations can be made on the anisotropic lattice. This
data will provide a useful forum for making comparisons with effective string 
theories and other models of the confining flux. 
Qualitative predictions of this fine structure remains a significant
challenge. Continuum extrapolations are needed and 
simulations with larger torelons are also under study.

This work was supported by the U.S.~NSF under Award PHY-0099450, U.S.~DOE 
Grant No.  DE-FG03-97ER40546, EU HPRN-CT-2000-00145. FM is grateful for 
support from Trinity College and Enterprise-Ireland.

\vspace{-1ex}

\begin{thebibliography}{9}
\bibitem{JKM1} J.K. Juge, J. Kuti, C. Morningstar, Phys.Rev.Lett.90:161601,2003.
\bibitem{JKM2} J.K. Juge, J. Kuti, C. Morningstar, Nucl.Phys.Proc.Suppl.106:691-693,2002
\bibitem{CM} C. Morningstar and J. Kuti, Confinement 2003, RIKEN.
\bibitem{Teper1} B. Lucini and M. Teper, Phys. Rev.D64:105019,2001.
\bibitem{PS} J. Polchinski and A. Strominger, Phys.Rev.
Lett.67:13,1991.
\end{thebibliography}
\end{document}